\documentclass[%
 aip,
 amsmath,amssymb,
 reprint,%
]{revtex4-1}
 \usepackage{amsmath, amsthm, amssymb,amsfonts}
\usepackage{graphicx}% Include figure files
\usepackage{dcolumn}% Align table columns on decimal point
\usepackage{bm}% bold math
\usepackage{color}
\usepackage[abs]{overpic}
\usepackage{multirow}
\usepackage{makecell}
\begin{document}

\title{Dispersive and kinetic effects on kinked Alfv\'en wave packets: a comparative study with fluid and hybrid models}

\author{Anna Tenerani}\email{Anna.Tenerani@austin.utexas.edu}

\affiliation{Department of Physics, University of Texas at Austin, TX 78712}
\author{Carlos Gonz\'alez}
\affiliation{Department of Physics, University of Texas at Austin, TX 78712}
  \email{Anna.Tenerani@austin.utexas.edu}
\author{Nikos Sioulas}
\affiliation{Department of Earth, Planetary and Space Sciences, University of California, Los Angeles, CA 90095}
\author{Chen Shi}
\affiliation{Department of Earth, Planetary and Space Sciences, University of California, Los Angeles, CA 90095}
\author{Marco Velli}
\affiliation{Department of Earth, Planetary and Space Sciences, University of California, Los Angeles, CA 90095}

\date{\today}

\begin{abstract}
 We investigate dispersive and kinetic effects on the evolution of a two-dimensional kinked Alfv\'en wave packet by comparing results from MHD, Hall-MHD and hybrid simulations of a low-$\beta$ plasma. We find that the Hall term determines the overall evolution of the wave packet over a characteristic time $\tau^*=\tau_a\ell/d_i$  in both fluid and hybrid models. Dispersion of the wave packet leads to the conversion of the wave energy into internal plasma energy.  When kinetic protons are considered, the proton internal energy increase has contributions from both plasma compressions and phase space mixing. The latter occurs in the direction parallel to the guiding mean magnetic field, due to protons resonating at the Alfv\'en speed with a compressible mode forced by the wave packet. Implications of our results for switchbacks observations and solar wind energetics are discussed. 
\end{abstract}

\maketitle

\section{introduction}

After E. Parker put forward the first theory of a supersonic solar wind~\citep{parker}, it has become clear that a thermally driven wind cannot reproduce the highest  speeds measured in-situ, reaching values up to $750-800$~km/~\citep{shi_2022, halekas_2022}, and that an additional source of energy is required to explain both the heating of the solar corona and the acceleration of high speed streams. Since the first observations in interplanetary space~\citep{coleman}, turbulence and waves have  been proposed  as a mechanism to heat and accelerate the solar wind plasma~\citep{cranmer_2015}. Alfv\'enic fluctuations indeed represent the dominant contribution to solar wind turbulence, especially, but not limited to, the fastest streams~\citep{damicis_2020}. Switchbacks are part of this turbulent flux continually emitted by the sun and correspond to large amplitude Alfv\'enic kinks of the magnetic field that lead to a local magnetic field polarity reversal. Parker Solar Probe (PSP) observations have shown that switchbacks are a common feature of the Alfv\'enic wind, regardless of its speed~\citep{horbury_2020, dudok_2020}.  

What is the origin of switchbacks and what is their role in solar wind dynamics and turbulence still remain important open questions. On the one hand, it has been proposed that switchbacks are an intrinsic part of the evolving turbulence and that they form in-situ dynamically, driven by solar wind  expansion or large scale shear flows~\citep{landi_2006,squire_2020, ruffolo_2020}. Alternatively, it has been argued that switchbacks are kinked magnetic field lines resulting from interchange reconnection in the corona. In the latter scenarios, switchbacks are considered as the outcome of processes that may generate the wind itself~\citep{schwadron_2021, zank_2020, drake_2020}. An analysis of the occurrence rate of switchbacks as a function of radial distance has shown that the probability to observe longer duration switchbacks increases with radial distance, while the probability to observe shorter duration switchbacks, shorter than a few tens of minutes, decreases. These observations suggest that the dynamics of switchbacks is complex and scale dependent, and that different types of switchbacks may coexists --- those formed closer to the corona and that gradually decay or degrade as they propagate out, and those formed in-situ as the turbulence evolves with radial distance~\citep{tenerani_2021}. 

An important property of Alfv\'enic fluctuations in the solar wind, including switchbacks, is that they are characterized by a nearly constant magnetic field amplitude~\citep{matteini_2015}. Large amplitude monochromatic or broadband magnetic and velocity fluctuations correlated like Alfv\'en waves are an exact solution --- although unstable --- to the nonlinear compressible Magnetohydrodynamic (MHD) model, provided the total magnetic field magnitude is also constant \citep{derby, malara_Phys_fluids_1996, tenerani_2013}. In previous work, we showed via MHD simulations that the parametric instability leads to the disruption of an Alfv\'en wave packet similar to a switchback over a timescale that can reach up to a few hundreds of Alfv\'en times~\citep{tenerani_2020}. Those results support the idea that some of the observed switchbacks may be formed back in the corona and then propagate out to distances of a few tens of solar radii, before  eventually decaying. On the other hand, switchbacks  occur over a wide range of scales. They can be as long as several hours (MHD scales), and as short as a few seconds, approaching the proton cyclotron period.  Switchbacks can therefore be affected also by dispersion and other kinetic effects faster than expansion and MHD processes, like the parametric decay instability mentioned above. Wave activity at the scale of a few seconds at switchback boundaries and degraded switchbacks with signatures of magnetic holes have been  reported during the first encounter of  PSP  \citep{farrell_2021}. Emission of dispersive waves can thus provide another channel for switchbacks' evolution and disruption. The goal of this work is to determine how and on what timescales dispersion and wave-particle interactions affect the dynamics of Alfv\'enic wave packets such as switchbacks. 

It is known that broadband Alfv\'enic fluctuations are no longer an exact solution to the Hall-MHD model due to dispersion. Exact nonlinear solutions to the Hall-MHD system still exist, but in the form of  monochromatic right or left-handed circularly polarized waves~\citep{sakai,hollweg}. Nonlinear Alfv\'enic solutions to reduced equations, namely, the DNLS (derivative nonlinear Schr{\" o}dinger) equation, have also been found in the form of one-dimensional solitary wave-packets with a modulated envelope~\citep{mjolus,spangler1,spangler2}. The effect of dispersion on large amplitude  Alfv\'enic fluctuations in plane geometry has been investigated extensively in the past~\citep{spangler_1985,hollweg, champeaux_1999, buti_2000, araneda_2008, matteini_2010, gonzalez_2021}  but, to the best of our knowledge, never for a two-dimensional wave packet such as the one considered here. 

In this work we consider a  two-dimensional Alfv\'en wave packet with constant magnetic and thermal pressure in quasi-parallel propagation and investigate  the role of dispersive and kinetic effects on its dynamics by comparing results from MHD, Hall-MHD and hybrid simulations. In section~\ref{numerical_models} we describe the numerical models used and the initial conditions for our simulations. Results are reported in Section~\ref{results}. We discuss the implications of our results on switchbacks lifetime and observations in Section~\ref{discussion} and we provide a summary in Section~\ref{summary}. 

\begin{table}[t]
\caption{Summary of the simulation runs, where $\ell_x$ is the longitudinal length of the wave packet, $d_i$ the proton inertial length, $\tau_a=B_{0x}/\sqrt{4\pi\rho_0}$ and $\tau^*=\tau_a \ell_x/d_i$.  }
\begin{center}
\begin{tabular}{ccccc}

	   & $(L_x\times L_y)/\ell_x$ &$\ell_x/d_i$ & $\beta$ & $\tau_*/\tau_a$\\
\hline
run 0 (MHD)& \multirow{4}{*}{$33.5\times8.37$} & $\infty$ &  \multirow{4}{*} {0.5}	&$\infty$\\

run 3 (Hall)&  & $150$ & 	&150\\

run 1 (Hall)&  & $30$ &	& 30\\

run 2a (Hall)&  & $6$ &	&6\\

run 2b (hybrid)& $34.13\times 8.53$ & $6$ & $\beta_{p,e}=0.25$ & 6\\
\hline
\end{tabular}
\end{center}
\label{table}
\end{table}%

\section{Numerical models}
\label{numerical_models}
We make use of a  2.5D numerical code (two-dimensional domain and three-dimensional vectors) that integrates the full set of compressible Hall-MHD equations in conservative form, where an adiabatic closure is assumed~\citep{shi_2019}. The code adopts periodic boundary conditions and derivatives are calculated via the Fast Fourier Transform. An explicit third-order Runge-Kutta method is used for the time integration and the Courant–Friedrichs–Lew condition is used to determine the appropriate time step. We do not impose explicit resistivity and use instead a pseudo-spectral filter to avoid energy accumulation at the grid-scale.

Simulations are initialized with a two-dimensional analytical model for a switchback as discussed in \citet{landi_2005}  and \citet{tenerani_2020}. The magnetic field is defined starting from the two-dimensional magnetic scalar potential $\psi(x,y)$,
\begin{equation}
{\bf B}=  {\boldsymbol\nabla}\times\psi(x,y){\bf \hat z}+B_z(x,y){\bf \hat z} + B_{x0}{\bf \hat x}, 
\label{equil}
\end{equation}
where 
\begin{equation}
\psi(x,y)= - \psi_0 \left( e^{-r_1^2}-e^{-r_2^2} \right),
\label{equil1}
\end{equation}
\begin{equation}
r_{1,2}^2=\left(\frac{x-x_{1,2}}{\ell_x}\right)^2+\left(\frac{y-y_{1,2}}{\ell_y}\right)^2.
\label{equil2}
\end{equation}

The component $B_z$ of the magnetic field is then determined by imposing a constant total magnetic field strength $B$,   
\begin{equation}
B_z(x,y)^2 = B^2-[B_x(x,y)^2+B_y(x,y)^2]. 
\label{equil3}
\end{equation}

The velocity fluctuation ${\delta\bf u}$ follows directly from the Alfv\'enicity condition, ${\delta\bf u}=-{\bf \delta B}/\sqrt{4\pi \rho_0}$,  ${\bf \delta B}$ being the fluctuating magnetic field~\citep{primavera_2019, tenerani_2020}.

The setup described above corresponds to a  wave-packet similar to a switchback localized in the $(x,y)$ plane, with initial constant magnetic field strength, that propagates in a homogeneous plasma with density $\rho_0$, pressure $p_0$, and guiding mean field ${\bf B}_{0x}$ that we take in the ${\bf\hat x}$ direction. Lengths are normalized to a reference length $L$, the magnetic field to $B_{0x}$, density to the background density $\rho_0$, speed to the corresponding  Alfv\'en speed.  In these units, the wave packet has length $\ell_x=1.5$, width $\ell_y=1$, and we set $|x_1-x_2|=|y_1-y_2|=2$. We have considered three Hall-MHD cases, each corresponding to a different value of the normalized proton inertial length, $d_i=0.25, 0.05, 0.01$ (where $d_i=v_a/\Omega_{ci}$, with $\Omega_{ci}=eB_{0x}/(m_i c)$ and $v_{a}=B_{x0}/\sqrt{4\pi\rho_0}$), and one MHD case ($d_i=0$).  Decreasing $d_i$ allows us to increase the scale separation between the typical length of the wave packet and dispersive scales. In all of the runs the plasma beta $\beta=8\pi p_0/B_{0x}^2$ is set to $\beta=0.5$ and the mesh resolution is $\Delta x=\Delta y=0.1$ with a domain size of  $L_x\times L_y=(8\times 2)2\pi$. The mesh resolution has been chosen to well resolve the dynamical evolution of system. A summary of the numerical parameters used is  reported in Table~\ref{table}.

The analytical form of the magnetic field given in eq.~(\ref{equil})-(\ref{equil3}) includes a finite mean magnetic field ${\bf B}_{0z}$ in the $z$-direction, which is required to maintain a constant magnetic field strength and whose magnitude depends on the amplitude of the fluctuation $\psi_0$. The mean field ${\bf B}_{0z}$ mimics self-consistently the Parker spiral magnetic field (at a fixed angle). We therefore consider a small but finite amplitude fluctuation, $\psi_0=0.25$, which allows us to simulate an Alfv\'enic wave-packet propagating at an angle $\theta=\arctan(B_{z0}/B_{x0})=0.61$ ($\theta\simeq 35^\circ$). The corresponding minimum value of the longitudinal field $B_x$ is $B_{x}^{(min)}=0.76$. A larger value for $\psi_0$  leading to a longitudinal field reversal with $B_x\lesssim 0$ would require a larger propagation angle, which strongly affects dispersion properties and would make a comparison with observations difficult. The amplitude of the fluctuations may also affect parametric instabilities. Here we have chosen the parameters so that the Alfv\'en wave-packet is  in quasi-parallel propagation and stable in the MHD limit. This allows us to focus solely on dispersive effects.  

 To investigate proton kinetic effects, we performed 2.5D (2D 3V for particles) simulations with the hybrid code CAMELIA (see, e.g., \citet{franci_2018}) in which electrons are described as a massless isothermal fluid. The hybrid code is periodic,  and it uses the current advance method~\cite{matthews} and Boris scheme for the particle pusher. Explicit resistivity ($\eta=0.001$ in units of $4\pi/c v_a\omega_{pi}^{-1}$, where $\omega_{pi}=\sqrt{4\pi n e^2/m_i}$) has been added to improve conservation of energy by avoiding the formation of magnetic fluctuations at the grid scale.  In the hybrid code, lengths are normalized to $d_i$ and time to the inverse of the proton gyrofrequency $\Omega_{ci}$. We impose the same initial condition as in the Hall-MHD simulations, an initial Maxwellian proton distribution and $\ell_x/d_i\simeq 6$, so that the hybrid simulation can be compared with run 2a. We use 2000 particles-per-cell, a mesh resolution of $\Delta x=\Delta y=0.2d_i$ and a domain size of $L_x\times L_y=(204.8\times51.2) d_i$.  In the hybrid simulation energy is conserved within 0.016\% and in the fluid code within $6\times 10^{-7}\%$. A summary of the numerical parameters is reported  in Table~\ref{table}. Both fluid and hybrid simulations are performed in a frame moving with the wave-packet at the Alfv\'en speed $v_{a}$. 

\section{Results} 
\label{results}
\subsection{Fluid model}

\begin{figure}
\includegraphics[width=0.5\textwidth]{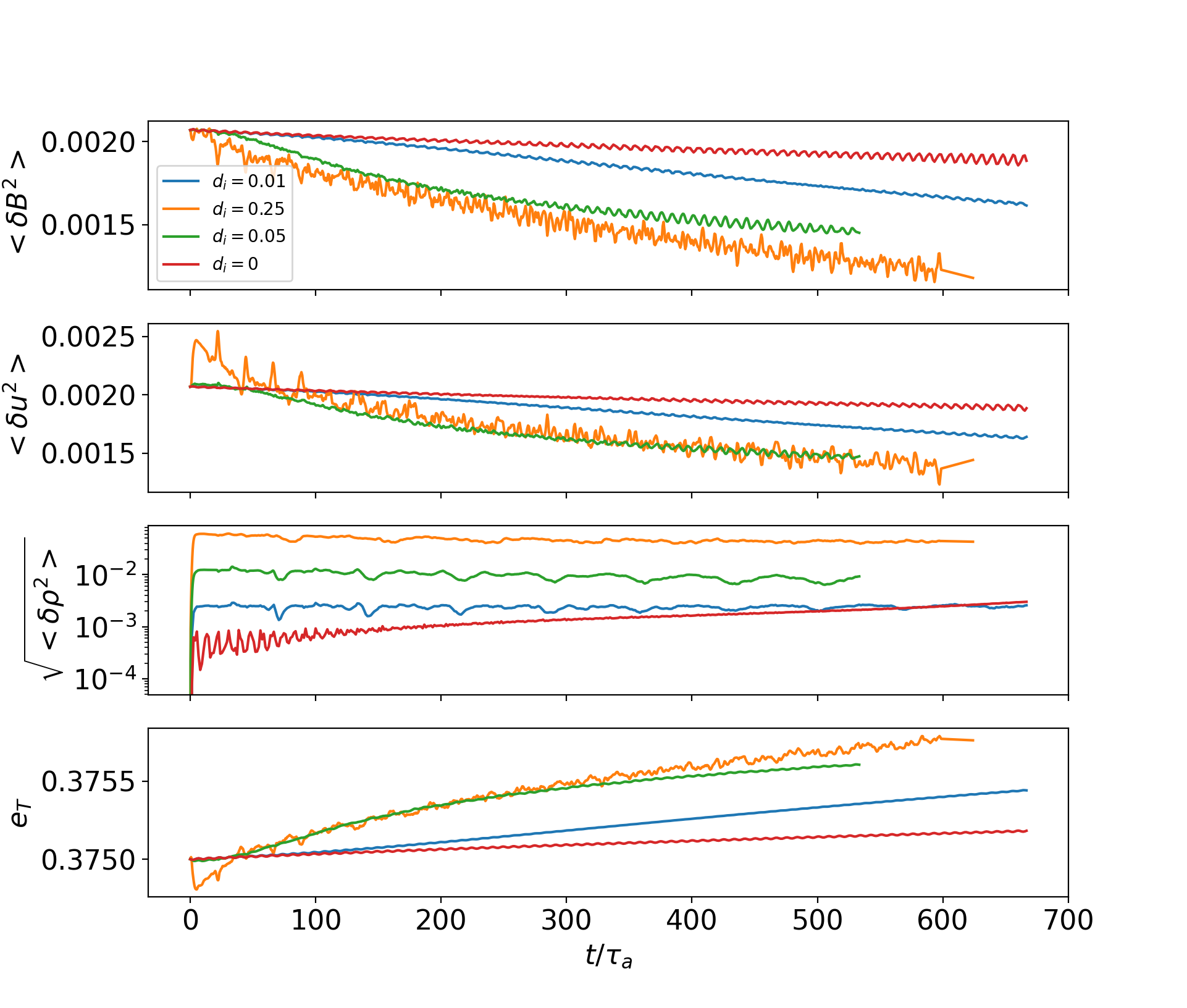}
\caption{MHD and Hall-MHD runs. Variance of magnetic  and velocity  fields (top and second panel),  rms of density fluctuations (third panel) and internal energy (bottom panel) as a function of time in units of Alfv\'en time.}
\label{overview1}
\end{figure}

\begin{figure}
\includegraphics[width=0.5\textwidth]{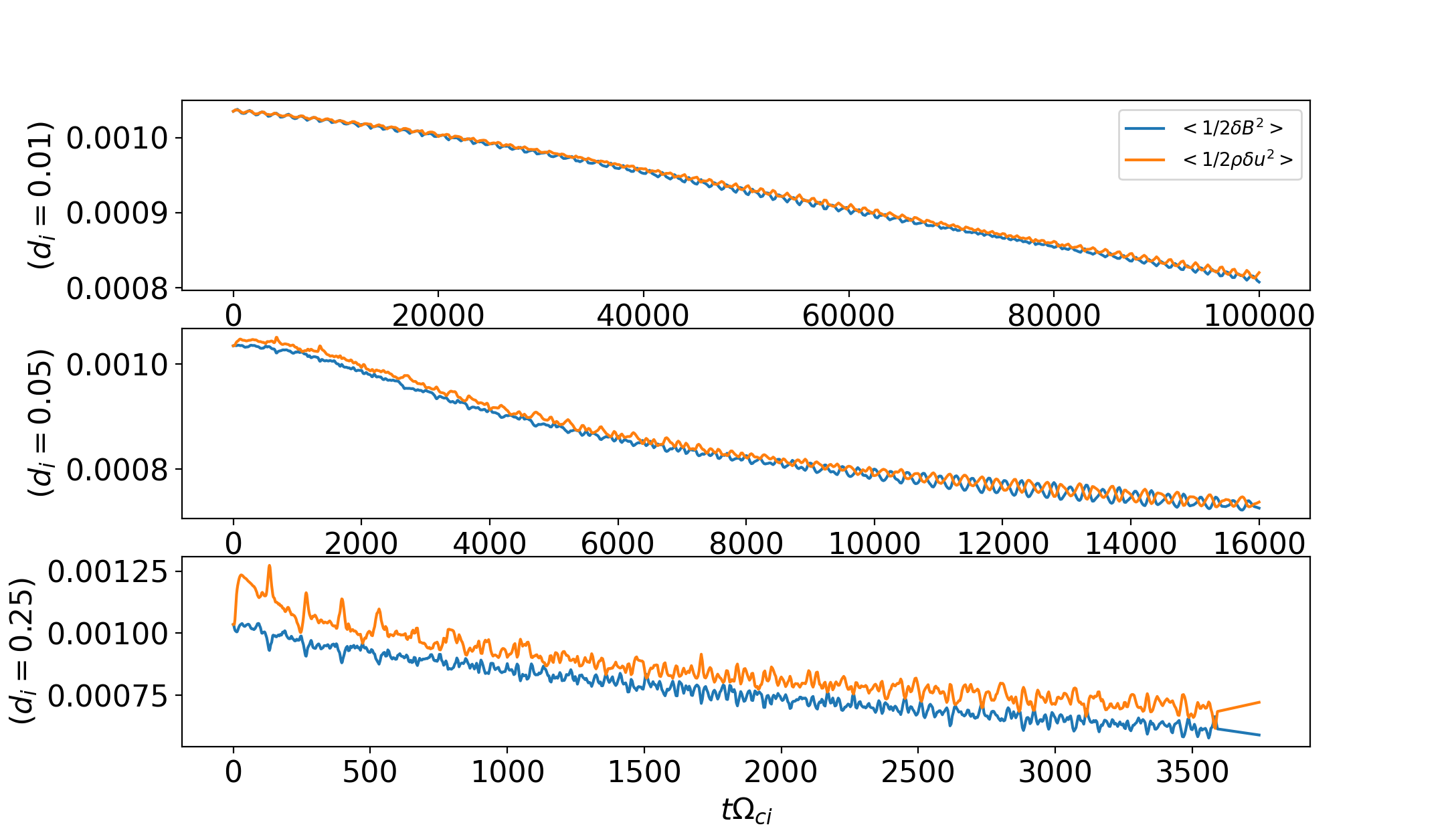}
\caption{Hall-MHD runs. Fluctuations' magnetic and kinetic energy as a function of time in units of  the inverse proton  gyrofrequency~$\Omega_{ci}$ for run 3 (top panel), run 1 (middle panel) and run 2a (bottom panel).}
\label{overview2}
\end{figure}

In Fig.~\ref{overview1} and Fig.~\ref{overview2} we report the overview of the evolution of the system for the Hall-MHD and MHD simulations.  Figure~\ref{overview1} displays the variances of the magnetic and velocity fields (top and second panels, respectively), the root-mean-square (rms) of density fluctuations (third panel) and the evolution of the internal energy (bottom panel) as a function of time in units of the Alfv\'en time $\tau_a=\ell_x/v_{a}$. The MHD simulation ($d_i=0$, red color) remains nearly stationary. The slight decrease in the variances of magnetic and velocity fields  is due to a continuous, slow growth of compressible fluctuations. Such a compressible fluctuations are expected to grow at a slow rate due to parametric instabilities, however, they remain negligible over the time interval considered, reaching a maximum value of $\delta \rho_{rms}/\rho_0=0.003$. The wave packet is thus not affected by the parametric instability in the MHD limit, at least until time $t=600\tau_a$. 

When dispersion is included,  a decrease in the kinetic and magnetic energy of the fluctuations is observed, together with an increase of the internal energy and of density fluctuations. In particular, the internal energy gain matches the loss in kinetic and magnetic energy of the wave packet, that is, 
\begin{equation}
\Delta( <1/2 \rho |\delta {\bf u}|^2+1/2|\delta {\bf B}|^2>)=-\Delta e_T,
\end{equation} 
where $\Delta g=g(t)-g(0)$ and the internal energy is $e_T=3/2<p>$. Conversion of magnetic and bulk kinetic energy  into internal energy is due to the coupling of the Alfv\'en wave packet to compressible modes, which becomes stronger as $\ell_x/d_i$ decreases.  As a consequence, the largest internal energy gain (for the time intervals considered) occurs in run 2a, where  the internal energy increases by $\Delta e_T/e_T(0)\simeq 0.2\%$ when the wave packet has undergone complete dispersion and has disrupted, with $\Delta <|\delta {\bf B}|^2>/<|\delta {\bf B}(0)|^2>\simeq -40\%$ and $\Delta <\rho|\delta {\bf u}|^2>/<\rho|\delta {\bf u}(0)|^2>\simeq -33\%$. Also the amplitude of density fluctuations  in run 2a are larger  than in the less dispersive cases, with a value of $\delta \rho_{rms}/\rho_0\simeq 0.04$. Although compressible fluctuations are generated in all of our simulations, the wave packet does not display signatures of instabilities, such as  modulational instability, even when dispersion is included. As it will be discussed later, the evolution of the wave packet is determined primarily by the Hall term in Ohm's law.

\begin{figure*}
\includegraphics[width=0.45\textwidth]{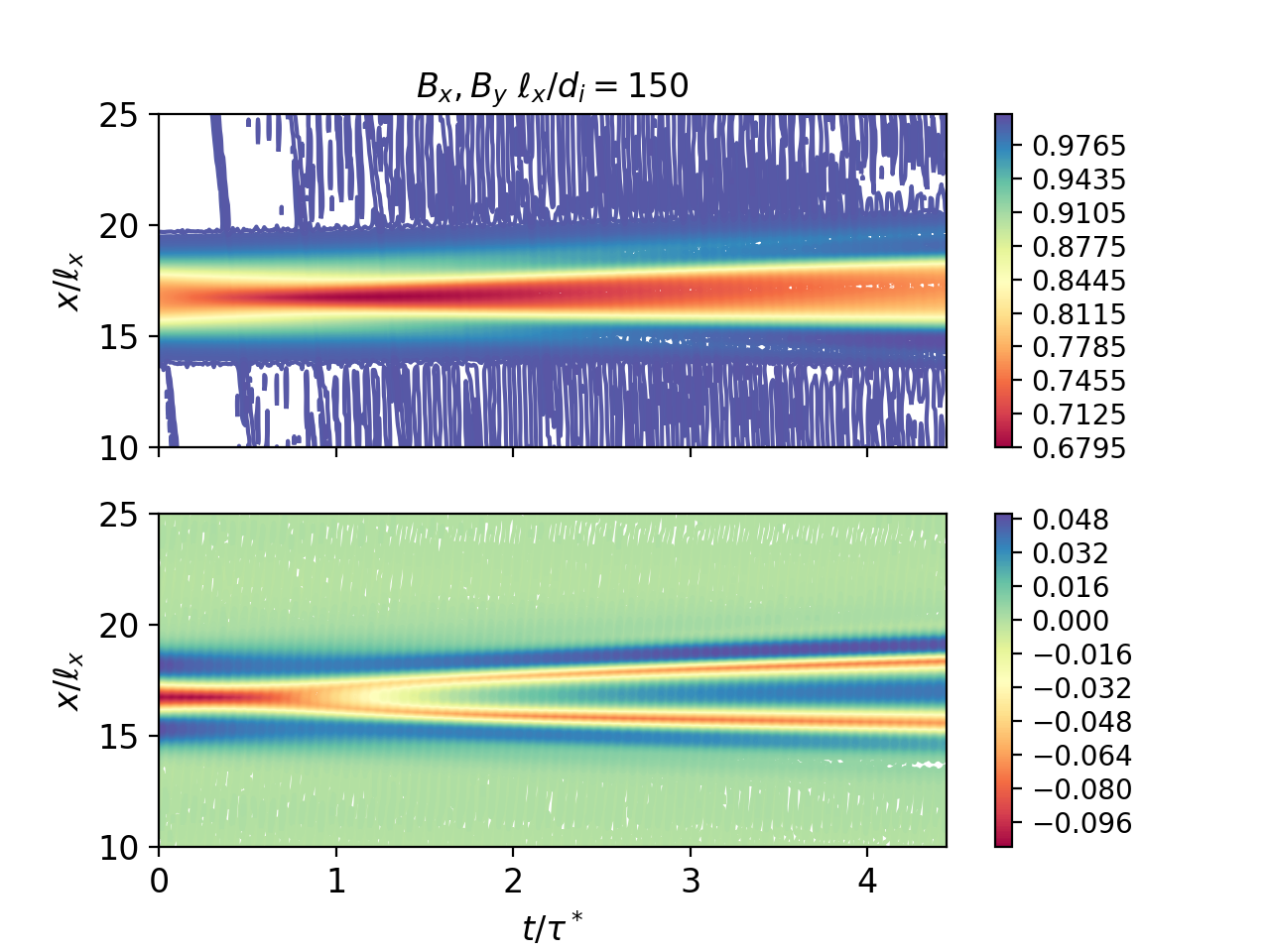}
\includegraphics[width=0.45\textwidth]{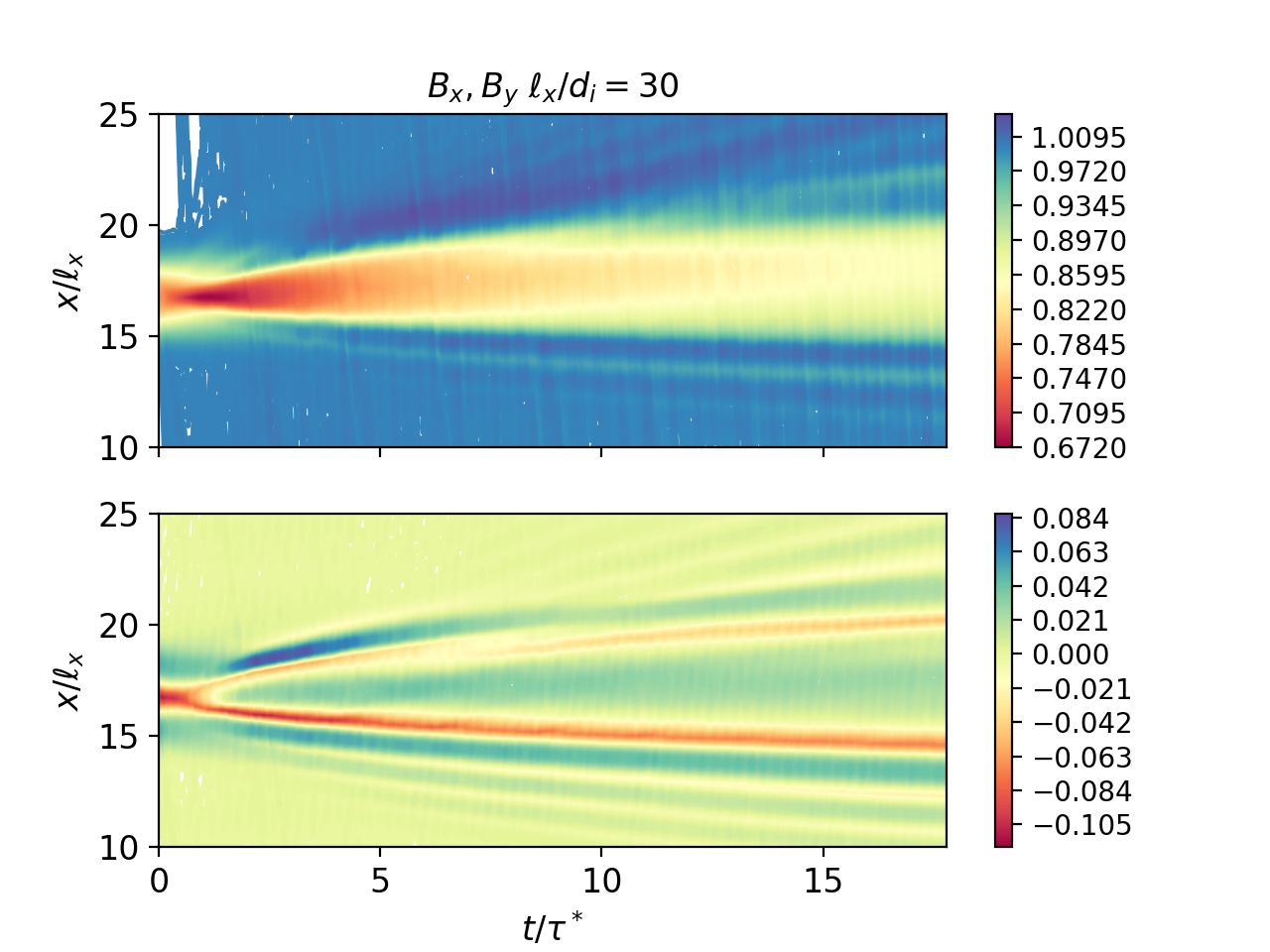}
\includegraphics[width=0.45\textwidth]{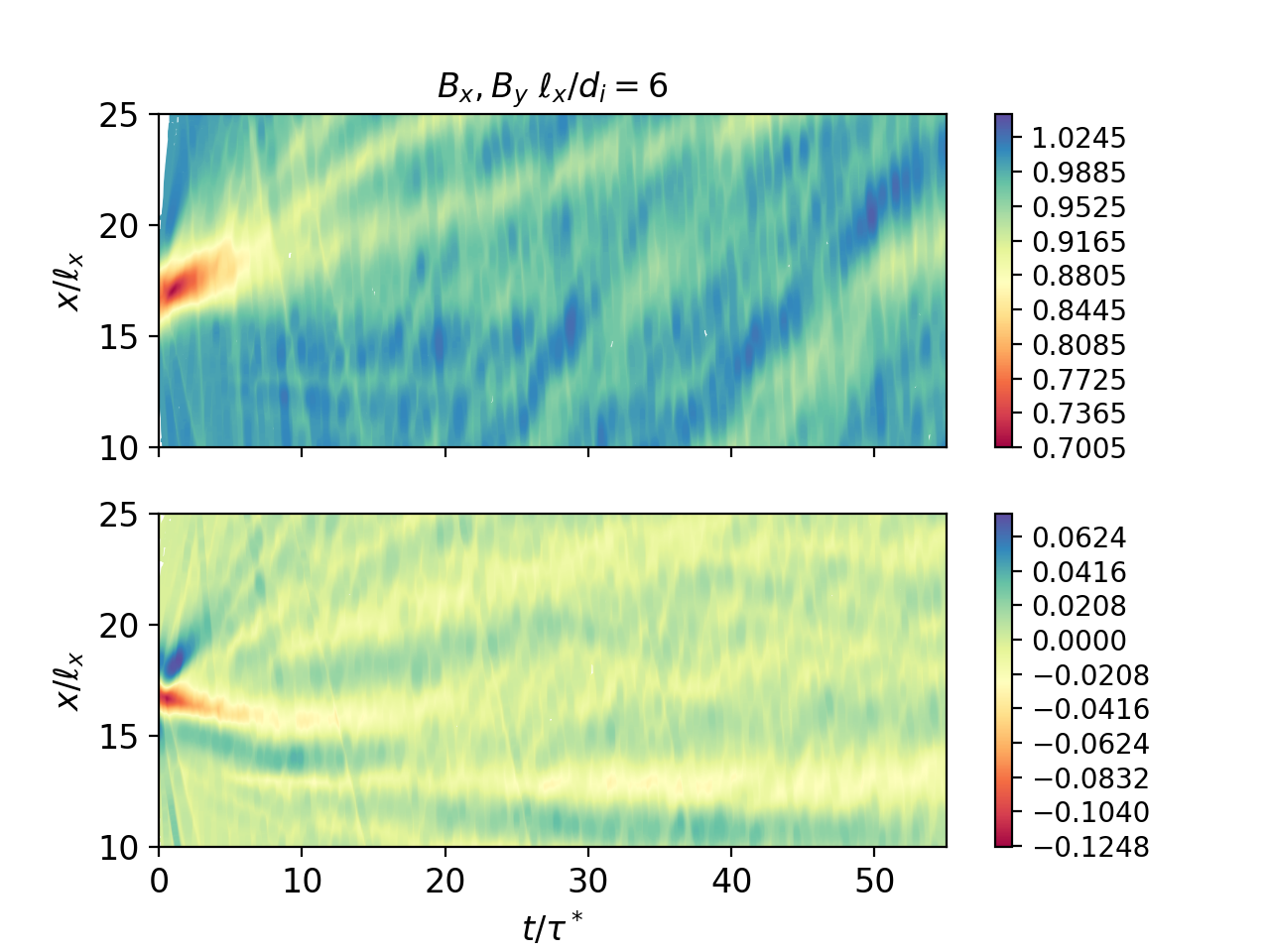}
\includegraphics[width=0.45\textwidth]{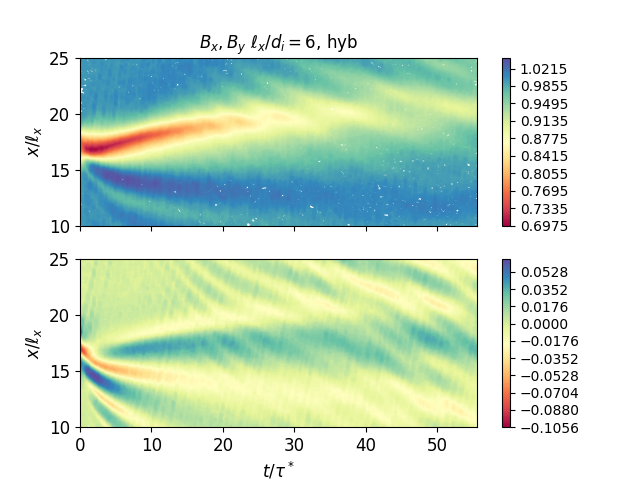}
\caption{Characteristic curves of the magnetic field  in the $(t,x)$ plane at $y=L_y/2$. Top panels: contour plot of $B_x$. Bottom panels: contour plot of $B_y$. From left to right, top to bottom, results for  run 3, run 1, run 2a and run 2b are shown.  Time is normalized to the characteristic time $\tau^*$ and length to $\ell_x$. Only a portion of the spatial domain is shown.}
\label{chars}
\end{figure*}

Figure~\ref{overview2} shows the evolution of the fluctuations' magnetic  and  kinetic energy density (blue and orange, respectively)  for the Hall-MHD simulations. We show, from top to bottom, results for run 3,  run 1 and run 2a as a function of time in units of $\Omega_{ci}^{-1}$. For very small but finite dispersion (run 3 and run 1) we find that magnetic and kinetic energy remain in equipartition to a very good approximation, while an excess of kinetic energy is observed in run 2a. In Fig.~\ref{chars} we show the characteristic curves  of the magnetic field components $B_x$ and $B_y$ in the $(t,x)$ plane at $y=L_y/2$, for run 3 (top left panels), run 1 (top right panels) and run 2a (bottom left panels). This set of simulations shows that the evolution of the wave packet is characterized by two stages marked by a characteristic time $\tau^*= \tau_a \ell_{x}/d_i$. The origin of the time $\tau^*$ will be discussed in section~\ref{model_hall}. 

\begin{figure}
\includegraphics[width=0.5\textwidth]{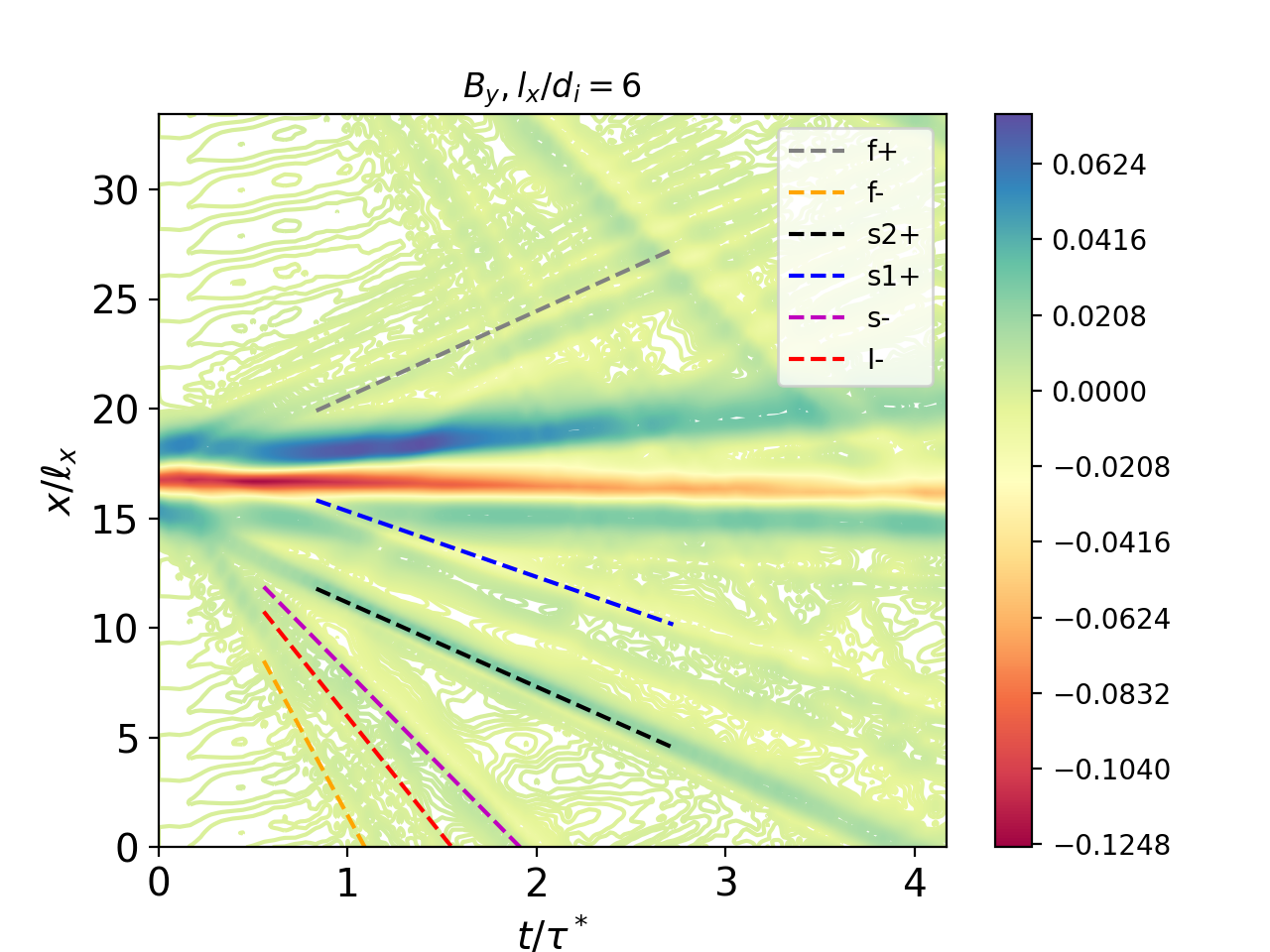}
\includegraphics[width=0.5\textwidth]{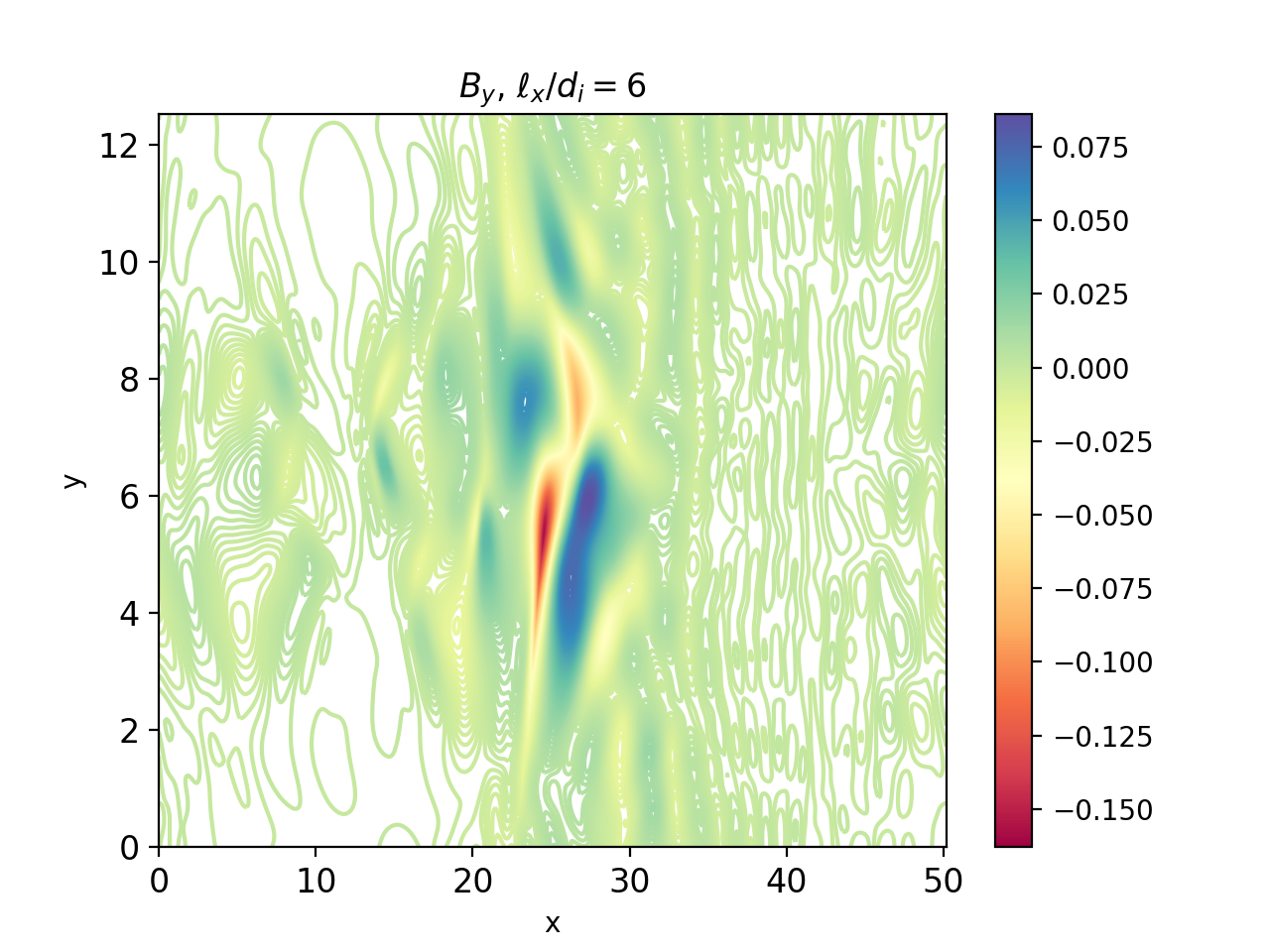}
\caption{Hall-MHD run 2a. Top panel: contour plot of $B_y$ in the $(t,x)$ plane at $y=L_y/2$. Dashed lines represent fast forward and backward (f+ and f$-$) modes, the backward intermediate (I-) mode and backward slow mode (s-). Two compressible wave packets of the slow mode type are also emitted and denoted as s1 and s2. Time is normalized to the characteristic time $\tau^*$ and length to $\ell_x$. Bottom panel: contour plot of $B_y$ in the $(x,y)$ plane at $t=1.4\tau^*$}
\label{chars_zoom_run2}
\end{figure}

During the first stage $t<\tau^*$, dispersive waves  that propagate ahead and behind the wave packet itself are emitted. Dispersive waves are found in all the simulations with dispersion, but their amplitude is negligible for weak dispersion (run 1 and run 3). We analize these emitted waves in Fig.~\ref{chars_zoom_run2}, top panel, where we show the same characteristic curves as in Fig.~\ref{chars} for run 2a, but for a shorter time interval. As a reference, we  also show the $(t,x)$ characteristic curves of fast, slow and intermediate modes that  match the emitted waves for this run. We have identified those modes by calculating the phase speed from the dispersion relation of the Hall-MHD system for $k d_i\lesssim1$. Although the system does not evolve through a nonlinear turbulent cascade, the magnetic energy spectrum (not reported here) shows that energy is transferred to that range of scales during the time interval $t\lesssim2.7 \tau^*$ (or $t\Omega_{ci}\lesssim 100$). In particular, the plotted curves correspond to phase speeds $v_f=\pm1.65$, $v_s=-0.46$ and $v_i=-0.8$ at $kd_i=0.87$ in the plasma rest frame.  In addition to dispersive waves, two slow-mode compressive wave packets localized in the $(x,y)$ plane are  emitted in all of the simulations with finite dispersion. Such a compressible wave packets correspond to localized structures, of about the size of the initial Alfv\'en wave packet, comprising magnetic pressure depletions and compressions anti-correlated with density fluctuations, propagating along the guiding field ($x$ direction) at nearly the slow mode speed (in the MHD limit the slow mode speed for our parameters is $v_s=0.5$). The two slow mode structures can be seen in Fig.~\ref{chars_zoom_run2}, top panel, in the form of a depletion and a bump of $B_y$, respectively,  moving at a speed of $v_{1}\simeq0.5$ (blue dashed line) and $v_{2}\simeq0.36$ (black dashed line) in the plasma rest frame. For the sake of illustration, Fig.~\ref{chars_zoom_run2}, bottom panel, shows the contour plot of $B_y$ in the $(x,y)$ plane at $t=1.4\tau^*$. In this plot, the dispersive waves are visible in the $(x,y)$ plane on the right of the wave-packet (which lies at the center); the slow modes correspond to the localized ``blobs'' lying on the left of the wave packet. Slow mode wave packets propagate at similar speeds also in run~1 and run 3,  for which we estimated $v_{1}\simeq0.5$ and $v_{2}\simeq0.45$ (not shown here).

The second stage starts  at  $t\simeq \tau^*$, when the initial Alfv\'en wave packet disperses by spreading out. During the dispersion stage  the kinetic and magnetic fluctuations remain in quasi energy partition, particularly for run 1 and run 3, as can be seen from Fig.~\ref{overview2}. Because of dispersion, large amplitude waves appear to propagate from the trailing and leading edge of the wave packet along the guiding field $B_{x0}$, as well as in the transverse ($y$) direction. Dispersion leads to a significant reduction of the longitudinal magnetic field  $\delta B_x$ on a timescale that ranges between $10\tau^*$ and $40\tau^*$. We find that in the higher dispersion case (run 2a) the longitudinal magnetic field has decreased by about 20\% with respect to its initial value at time $t=10\tau^*$, after which it remains stationary. Not surprisingly, when dispersion is weak (run 1 and run 3) and energy is not lost into many wave modes, the same 20\% decrease occurs later, at time $t=40\tau^*$, and the longitudinal fluctuation continues to slowly decrease making the wave packet shallower with time. 

In Section~\ref{model_hall} we discuss a simple model to interpret the evolution of the magnetic field over long time scales when compressibility is negligible.

\subsection{Hybrid model}

\begin{figure}
\includegraphics[width=0.5\textwidth]{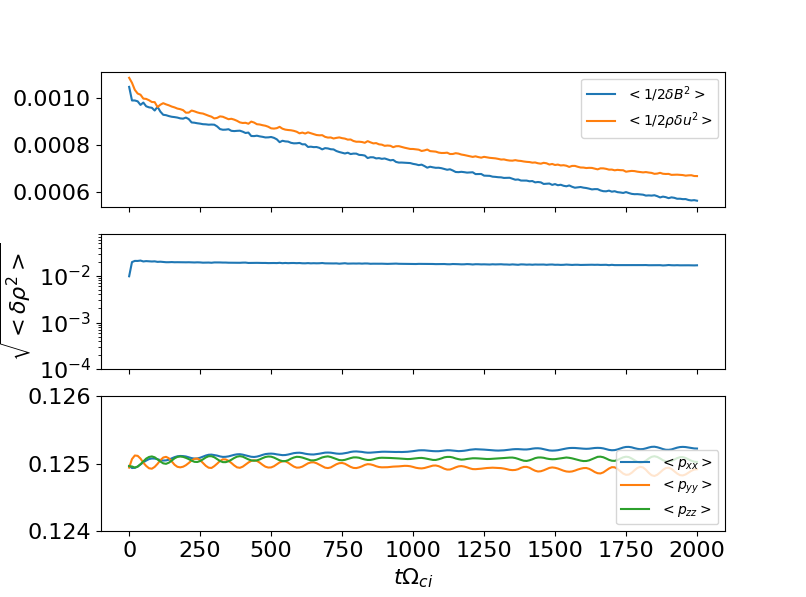}
\caption{Hybrid run 2b: time evolution of the magnetic and kinetic energy density of fluctuations (top panel), of the density rms (middle panel) and the diagonal mean pressures (bottom panel). }
\label{overview_hyb}
\end{figure}

In Fig.~\ref{overview_hyb} we show the overview of the time evolution of fluctuation's kinetic and magnetic energy density, of the density rms, and of the mean pressures for the hybrid run. The system evolves in a way which is consistent with the Hall-MHD runs shown in Fig.~\ref{overview1} and Fig.~\ref{overview2}. In Fig.~\ref{chars}, bottom right panel, we show the characteristic curves  of the magnetic field in the $(t,x)$ plane at $y=L_y/2$ for a comparison with the Hall-MHD model.  

\begin{figure}
\includegraphics[width=0.5\textwidth]{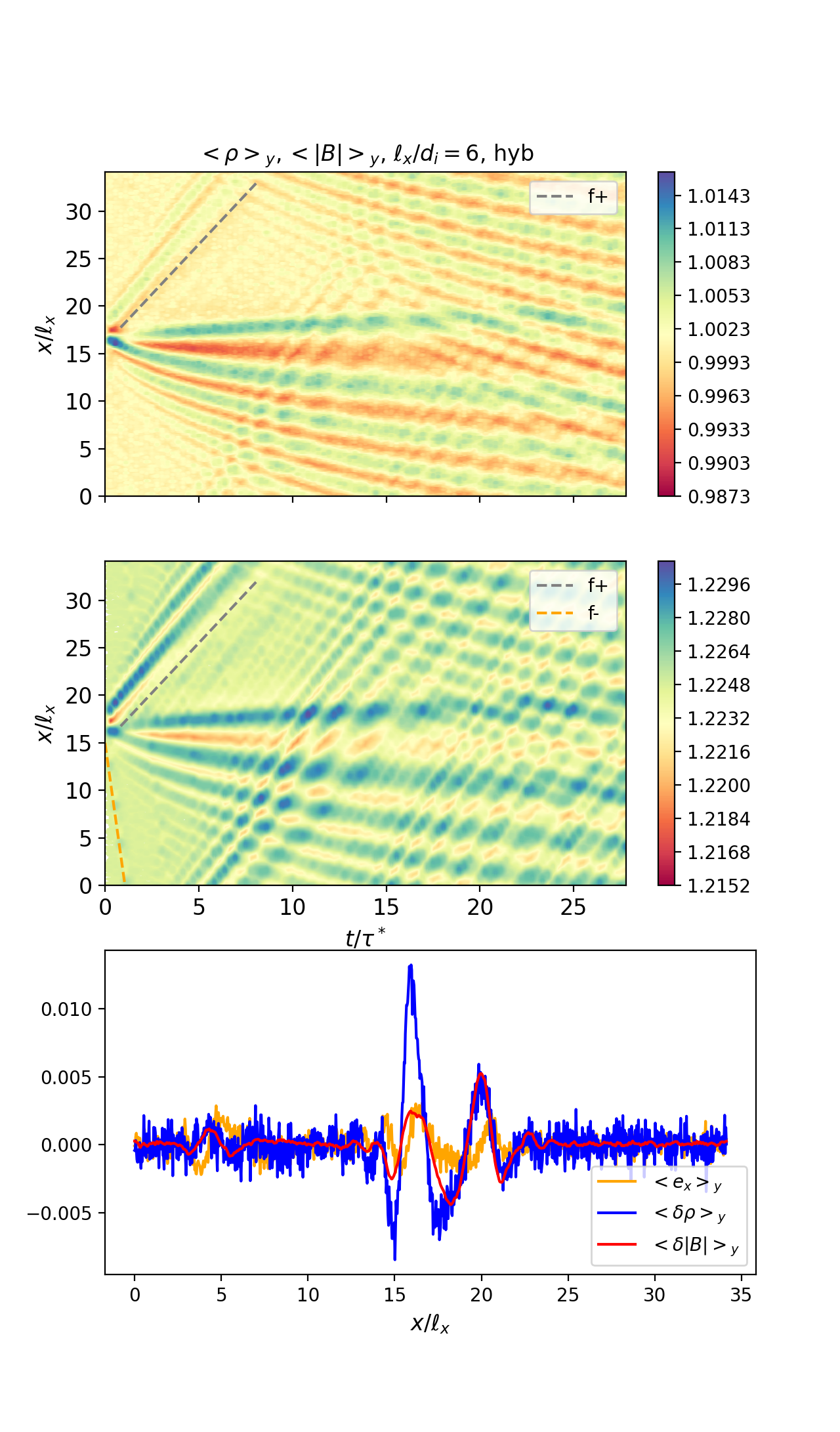}
\caption{Hybrid run. Top two panels: characteristic curves  of $<\rho(x,t)>_y$ and $<|{\bf B}(x,t)|>_y$, averaged over $y$.  Dashed lines indicate the forward and backward fast modes. Bottom panel: fluctuation of the averaged density and magnetic field strength, $<\delta \rho(x)>_y$ and $<\delta|{\bf B}(x)|>_y$, respectively, and of the electric field $<e_x(x)>_y$ at time $t\Omega_{ci}=30$ ($t=0.83\tau^*$).}
\label{hyb_chars_bis}
\end{figure}

Although the hybrid simulation is qualitatively similar to the Hall-MHD simulations, there are a few differences. In the hybrid model we do not observe slow-mode wave packets emitted at early times $t<\tau^*$, and density fluctuations are overall weaker, with $\delta\rho_{rms}/\rho_0\simeq0.018$. Fast modes with positively correlated density and magnetic pressure perturbations are instead emitted within a few tens of gyroperiods. Such a fast mode perturbations are generated within the Alfv\'en wave packet, forced by magnetic pressure imbalance induced by dispersive effects. Since the compressible mode  is forced, a portion of it remains stuck to the Alfv\'en wave packet propagating at the Alfv\'en speed, and a portion of it propagates ahead and behind the wave packet at a velocity of about $v_f\simeq\pm1.35 v_a$ (in the plasma rest frame). The fast modes can be seen in the top two panels of Fig.~\ref{hyb_chars_bis}, where we show the contour plots of $<\rho(x,t)>_y$ and $<|{\bf B}(x,t)|>_y$, averaged over the $y$ coordinate to reduce  noise in the density. The forward fast mode is clearly visible in both density and magnetic field magnitude contour plots. The backward fast mode is of smaller amplitude and is not  visible in the density contour. The fast mode perturbations and the corresponding signature in the electric field are  shown in the third panel of Fig.~\ref{hyb_chars_bis}, where we plot the fluctuation of density and magnetic field strength, $<\delta \rho(x)>_y$ and $<\delta|{\bf B}(x)|>_y$, respectively, and of the electric field $<e_x(x)>_y$  as a function of $x$ at time $t\Omega_{ci}=30$ ($t=0.83\tau^*$).

After time $t=\tau^*$, the evolution is consistent with the Hall-MHD simulations. However, in the hybrid model  $\delta B_x$ persists  longer than in its Hall-MHD counterpart (cfr. Fig.~\ref{chars}, bottom right panels, which correspond to the hybrid simulation, and bottom left panels, which correspond to the Hall-MHD run 2a). We ascribe the persistence of the wave packet over longer timescales  to the fact the in the hybrid model less energy is lost initially into different types of dispersive waves. 

\begin{figure}
\includegraphics[width=0.5\textwidth]{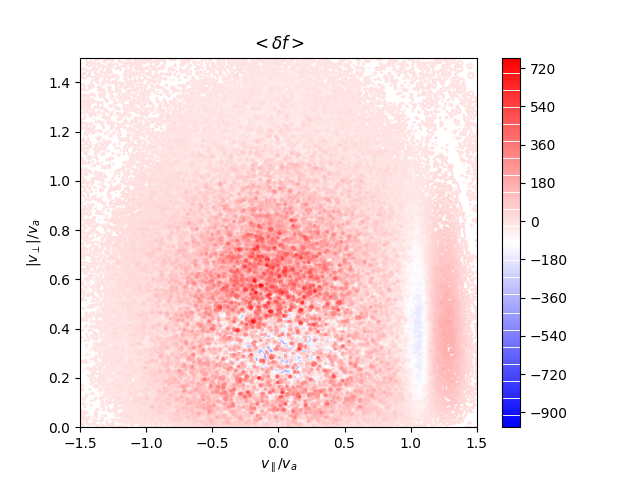}
\includegraphics[width=0.5\textwidth]{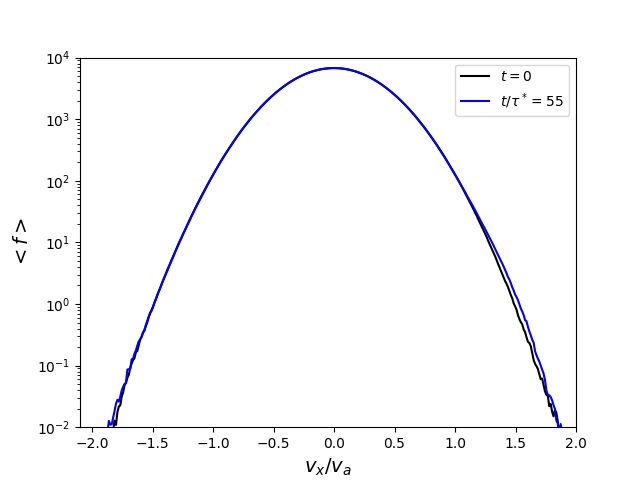}
\caption{Hybrid run. Top panel: contour plot of~$<\delta f({\bf x},v_\parallel,v_\perp)>_{\bf x}$ at $t\Omega_{ci}=2000$. Bottom panel: averaged distribution function $<f({\bf x}, {\bf v}, t)>_{{\bf x}, v_y, v_z}$ as a function of $v_x$ at $t=0$ and $t\Omega_{ci}=2000$ ($t=55\tau^*$).}
\label{dist_func}
\end{figure}

The field-aligned mean proton pressures, $<p_{xx}>$ and $<p_{zz}>$ shown in the bottom panel of Fig.~\ref{overview_hyb}, tend to increase as the wave packet  disperses, just like in the fluid simulations. The increase in proton internal energy is determined  by $<p_{xx}>$, which changes by $\Delta p_{xx}/p_{xx}(0)=0.2\%$ in a time interval $\Delta t=2000\Omega_{ci}^{-1}$, comparable to the relative change of internal energy in run 2a in the same time interval (which is $\Delta e_T/e_T(0)\simeq 0.14\%$). While in the fluid models the gain in internal energy is due to compressions of the plasma, in the hybrid model contributions from both compressions and phase space mixing determine the changes in internal energy of protons. The corresponding signature of proton heating in phase space is  reported in Fig.~\ref{dist_func}. The top panel shows the variation $\delta f$ of the spatially averaged proton distribution function $<\delta f({\bf x},v_\parallel,v_\perp, t)>_{\bf x}$ at $t\Omega_{ci}=2000$ ($t=55\tau^*$). The bottom panel shows the averaged distribution function $<f({\bf x},  {\bf v}, t)>_{{\bf x}, v_y, v_z}$ as a function of $v_x$ at $t=0$ and at $t\Omega_{ci}=2000$. It is interesting to see that $<\delta f>$ displays a clear structure in velocity space at the Alfv\'en speed, resulting in the small shoulder in $<f(v_x)>$ at $v_x>v_a$. Such a signature starts to emerge in phase space at time $t\Omega_{ci}=50$ ($t=1.38\tau^*$). At that time $<p_{xx}>$ has reached its highest increase rate, and density rms and electric field their highest decrease rate. This suggests that a wave-particle resonance exists with the  fast mode forced initially and discussed above. Proton resonance is however not sufficient to damp entirely density fluctuations.  In Sec.~\ref{discussion} we provide a rough estimate to compare the increase of internal energy found in our simulations with the one inferred from observations.

\section{Discussion}
\label{discussion}

\subsection{The Hall effect on Alfv\'en wave packets}
\label{model_hall}
\begin{figure}
\includegraphics[width=0.5\textwidth]{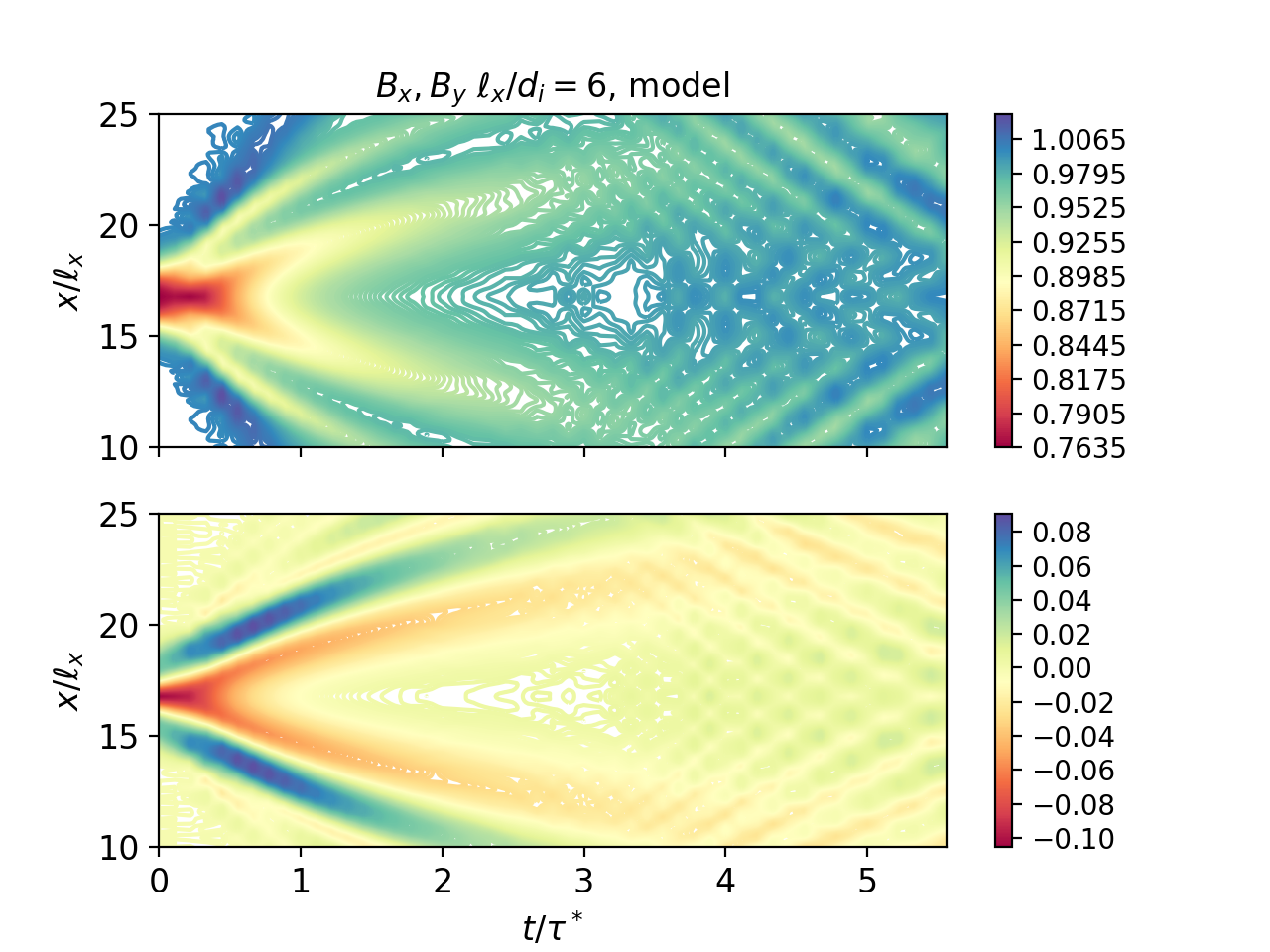}
\caption{Solution to eq.~(\ref{psi})-(\ref{bz}). Top panel: contour plot of $B_x(x,L_y/2,t)$. Bottom panel: contour plot of $B_y(x,L_y/2,t)$. Only a portion of the spatial domain is shown.}
\label{model}
\end{figure}

In order to interpret our simulation results, let us assume  that Alfv\'enicity ($\delta {\bf u}=\pm\delta {\bf B}/\sqrt{4\pi \rho_0}$) is conserved and that compressibility (thermal and magnetic) remains negligible for weak dispersion, two conditions that are satisfied in run 1 and run 3. In this case,  the motional electric field contributes to the propagation of the wave packet at the Alfv\'en speed, and departures from the MHD exact solution are determined only by the Hall term in Ohm's law. By neglecting the ${\bf u}\times {\bf B}$ term while retaining the Hall term,  the induction equation in a two-dimensional system can be cast into the following set of equations for the magnetic potential $\psi$ and the out-of-plane component $B_z$ of the magnetic field:

\begin{equation}
\frac{\partial \psi}{\partial t}=-d_i {\bf B}\cdot \boldsymbol\nabla B_z
\label{psi}
\end{equation}
\begin{equation}
\frac{\partial B_z}{\partial t}=d_i{\bf B}\cdot  \boldsymbol\nabla\nabla^2\psi, 
\label{bz}
\end{equation}

where we have normalized the magnetic field to  $B_{x0}$, lengths to an arbitrary length $L$, density to a reference $\rho_0$, and we have  approximated $\rho\simeq \rho_0$. Equations~(\ref{psi})-(\ref{bz}) introduce a characteristic time $\tau^*$. If the wave packet is nearly isotropic, as in the case considered here, $\tau^*\sim\tau_a\ell_x/d_i$. If the wave-packet is highly anisotropic then the timescale will be determined by the shortest time $\tau_{x,y}^*\sim\tau_a\ell_{x,y}/d_i$. When equations (\ref{psi})-(\ref{bz}) are linearized, they can be recast into an harmonic oscillator equation in Fourier space for the Fourier components $\tilde \psi(t, {\bf k})$, 

\begin{equation}
\frac{\partial^2}{\partial t^2} \tilde\psi(t,{\bf k})=-\omega_h^2 \tilde\psi(t,{\bf k}), 
\label{psi_lin}
\end{equation}

with $\omega_h^2=d_i^2|k_x|^2|{\bf k}|^2$ (in our normalization).  The amplitude of each Fourier mode is a periodic function of time, with a periodicity that depends on ${\bf k}$ and $d_i$. In general, the solution for a non-monochromatic wave packet  can be found  for given initial conditions $\psi(x,y,0)$ and $\partial\psi(x,y,0)/\partial t$, and by transforming back from Fourier to real space. We have solved eq.~(\ref{psi})-(\ref{bz}) with a 3rd order Runge Kutta scheme and pseudo-spectral method and eq.~(\ref{psi_lin}) by using eq.~(\ref{equil1})-(\ref{equil3}) as initial conditions, and we did not find significant quantitative differences. In Fig.~\ref{model} we show the  nonlinear solution to eq.~(\ref{psi})-(\ref{bz}) for $d_i=0.25$. Note that changing $d_i$ introduces a rescaling of time included into $\tau^*$, so that the plot in Fig.~\ref{model} can be compared with run 1 and run 3 as well (Fig.~\ref{chars}). As can be seen, the model reproduces quite well the observed evolution for run 1 and run 3, although the full Hall-MHD system evolves somewhat slower, by about a factor of two. Nevertheless, the main features of the long-term evolution, in particular wave dispersion along and across the guiding field that characterizes all of our simulations, can be recognized in this simple model and can therefore be ascribed to the Hall effect. This model does not capture entirely the evolution of run 2a and run 2b, where departures from equipartition of magnetic and kinetic energy are larger, and where compressible effects are in general more important.

In summary, the role of the Hall effect is twofold. When dispersion is weak, $\ell/d_i\gg 1$, the Hall term determines a slow dispersion of the initial wave packet that starts to affect its dynamics after a characteristic time~$\tau^*$. When dispersion becomes stronger,  $\ell/d_i\gtrsim 1$, departures from  MHD  are larger and thus the initial wave packet couples with compressible and other dispersive modes at times~$t<\tau^*$,  before dispersing at  about $t\simeq\tau^*$.   Such a coupling with compressible modes, mediated by dispersion, leads to an increase of internal energy.

Our results are remarkably different from Hall-MHD and hybrid simulations of 1D  Alfv\'en waves (broadband or monochromatic) and solitons. Previous work has shown that dispersion can cause modulational instabilities and wave steepening and collapse when a plane geometry is adopted for the fluctuation~\citep{spangler_1985, buti_2000, matteini_2010, gonzalez_2021}. Our work instead shows that the evolution of a 2D (non-plane) wave packet, in quasi-parallel propagation, differs significantly from its 1D counterpart (at least without initial strong amplitude modulations). First, just like without dispersion, a localized wave packet tends to be more stable than an equivalent large amplitude plane wave. Second, the two-dimensional dynamics allows for additional channels for wave evolution, inhibiting strong field aligned wave steepening. As a result, the overall evolution appears to be determined mainly by the Hall electric field leading to dispersion along and across the mean field. A field aligned beam, a signature of  steepened Alfv\'en wave fronts~\citep{matteini_2010, gonzalez_2021}, does not form in the case considered here. Instead, a wave-particle resonance is triggered by forced compressible modes, leading to parallel heating. 

We conclude by comparing the rate of change of the internal energy (either parallel or mean internal energy) resulting from the coupling with compressible modes with the rates of internal energy change estimated from observations \citep{hellinger_2011}. In particular, observations show that parallel pressure  increases with radial distance after $R\gtrsim 0.6$~au. We therefore provide an order of magnitude estimate of the heating rates by taking typical values of density and temperature at $R\simeq 0.6$ au. By using results from run 2b we can estimate $\Delta p_{xx}/\Delta t\simeq p_{xx}(0)\times 0.002/(2000\Omega_{ci}^{-1})$. With a number density of $n\simeq7.7\times 10^{6}\,m^{-3}$ at $R=0.6$~au, parallel temperature $T_\parallel\simeq 355764\,K$, and $2\pi/\Omega_{ci}\simeq 1\,s$, we obtain $\Delta p_{xx}/\Delta t\simeq 2\times10^{-16}\,W m^{-3}$. Such a value is of the same order of magnitude of the  parallel heating rate extrapolated from in-situ data at $R\geq 0.6-0.7$~au~\citep{hellinger_2011}.  A similar estimate for the total internal energy increase rate from run 2a, with $T\simeq 364845\,K$ at $R=0.6$~au, gives $\Delta e_T/\Delta t\simeq 4\times10^{-16}\,W m^{-3}$, still of the same order of the observational extrapolation.  This estimate provides a lower limit, in the sense that a larger amplitude wave packet can transfer more of its energy to internal energy. While this comparison is encouraging and suggests that dispersion of large amplitude wave packets can provide the required heating rates, perpendicular heating, which represents the  dominant contribution to the solar wind heating rate, in particular for $R<0.6$ au~\citep{hellinger_2011}, is not  observed in our simulations. Dispersion in the transverse direction of the wave packet however may impact the turbulent cascade and thus, indirectly, perpendicular heating when a fully turbulent plasma is considered. 

%both parallel heating and overall heating from a single wave packet is weak, in the sense that the resulting average rate of internal energy increase is much smaller than what estimated from solar wind data. By using results from run 2b we can estimate $\Delta p_{xx}/\Delta t\simeq p_{xx}(0)\times 0.002/(2000\Omega_{ci}^{-1})$. With a number density of $n\simeq7.7\times 10^{6}\,m^{-3}$ at $R=0.6$~au, parallel temperature $T_\parallel\simeq 355764\,K$, and $2\pi/\Omega_{ci}\simeq 1\,s$, we obtain $\Delta p_{xx}/\Delta t\simeq 6\times10^{-18}\,W m^{-3}$. This is about an order of magnitude less than the  parallel heating rate extrapolated from in-situ data at $R\geq 0.6$~au~\citep{hellinger_2011}.  A similar estimate for the total internal energy increase rate from run 2a, with $T\simeq 364845\,K$ at $R=0.6$~au, gives $\Delta e_T/\Delta t\simeq 10^{-17}\,W m^{-3}$, still an order of magnitude smaller than observational extrapolation. On the other hand perpendicular heating, which represents the  dominant contribution to the solar wind heating rate, is not  observed in our simulations. Dispersion in the transverse direction of the wave packet however  may impact the turbulent cascade and thus, indirectly, perpendicular heating when a fully turbulent plasma is considered.   

\subsection{Implications for switchbacks observations}

\begin{figure}
\includegraphics[width=0.5\textwidth]{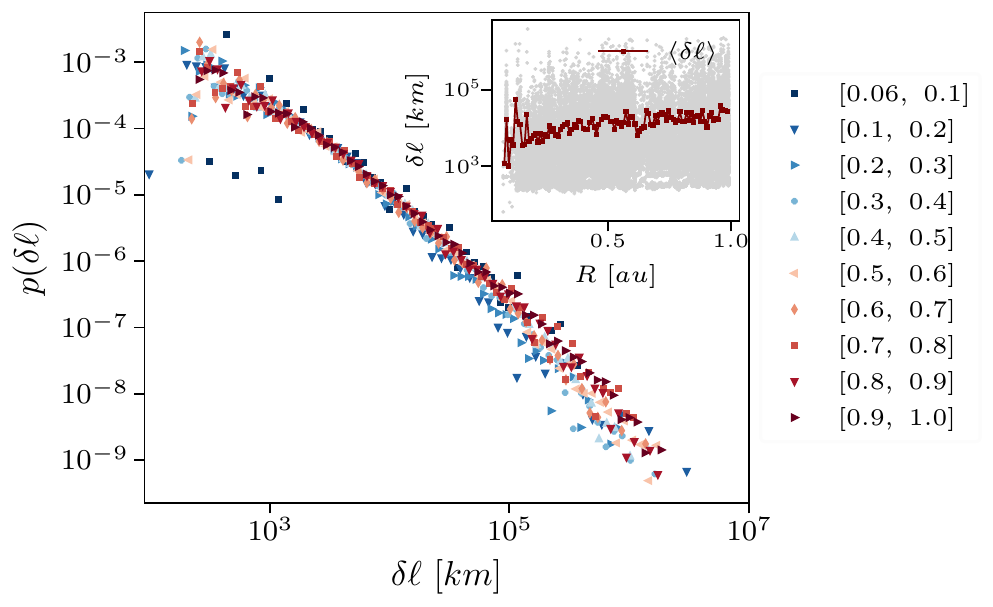}
\includegraphics[width=0.5\textwidth]{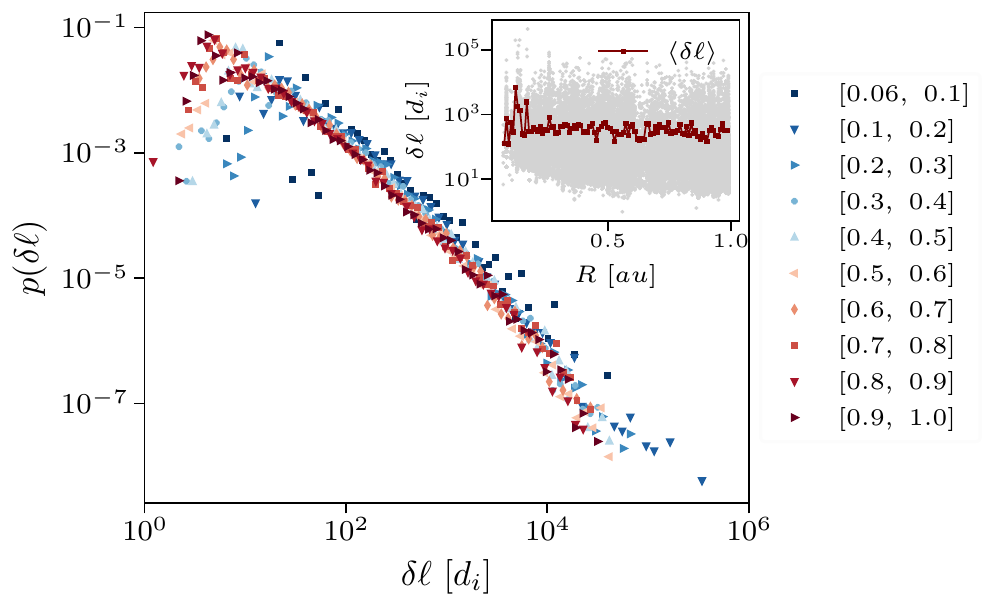}
\caption{Probability distribution of the length of switchbacks at radial distances $0.06<R<1$~au in units of km (top panel) or ion inertial length (bottom panel). The inset plots show the scatterplot of $\ell$ and $\ell/d_i$, respectively, as a function of $R$ (in gray), and the mean value for each radial bin is shown in~red.}
\label{data}
\end{figure}

Even if switchbacks are generally within the inertial range of scales, dispersive effects can still be non negligible for switchbacks at smaller scales ($\ell_{x,y}/d_i\gtrsim 1$), or, as our results show, affect the evolution of those at  larger scales ($\ell_{x,y} d_i\gg1$) over sufficiently long times. To provide context, we show in Fig.~\ref{data}  the probability distribution  of the length of switchbacks $p(\ell)$, where $\ell$ is expressed in units of km (top panel) and of the ion inertial length (bottom panel), for distances from the sun $R=0.06-1$ au. Details on the datasets and methods can be found in the Appendix. Our simulations cover the first decade of the distribution shown in the bottom panel of Fig.~\ref{data}. 
%As can be seen, the average length of switchbacks tends to increase with $R$, in agreement with our previous study. The radial trend of the mean of $\ell/d_i$ appears to be independent from $R$, although this may be a result of the fact that $d_i\propto R$ while the data resolution remains fixed. As a consequence, a portion of the distribution of $\ell/d_i$ towards the smaller values may be lost at the larger values of $R$. 

In prior work we determined that, in MHD, parametric decay can take a time of up to several hundreds of Alfv\'en times before destroying the switchbacks, if the wind is ``quiet'' (large system size)~\citep{tenerani_2020}. Here we have shown that if a wave packet is stable with respect to parametric instabilities, then dispersive effects determine the  time evolution of the wave packet. For example, if a switchback is stable over a few hundreds of Alfv\'en times, we expect the dispersive timescale $\tau^*$ to be shorter than that of parametric instabilities for those switchbacks of size $\ell/d_i\lesssim 100$. Switchbacks of duration $\delta t$ in the range $\delta t\simeq 10-100$~s in a wind with speed $V_{sw}=500$~km/s have an approximate size of $\ell\simeq (0.5-5)\times10^4$~km and $\ell/d_i\simeq (0.5-5)\times 100$ at a radial distance of $R\simeq0.1-0.2$~au (cfr. Fig.~\ref{data}). With a mean Alfv\'en speed of about $v_a\simeq 50$~km/s we obtain $\tau^*\simeq (5-500)\times 10^3$~s, independent from radial distance. Dispersive effects should start to degrade those switchbacks over a distance of $\Delta R\simeq V_{sw}\tau^*$. By taking $V_{sw}\simeq 500$~km/s, we obtain a corresponding range of $\Delta R\simeq 0.017-1.67$~au. Note that the large variation in $\Delta R$ is due to the fact that $\tau^*\propto\ell^2$. Nevertheless, this estimate indicates that there should be a subset of switchbacks undergoing dispersion within 1 au. 

Our kinetic and weakly dispersive fluid simulations show that while waves are emitted from the leading and trailing edge of the wave packet, causing a fast dispersion of the transverse components (i.e., tangential and normal in RTN coordinates), the longitudinal (i.e., radial) perturbation persists for several tens of $\tau^*$. Thus, dispersive effects on switchbacks could result in a defined field reversal in the radial direction with wave activity at its boundaries, including strong perturbations in magnetic pressure, and associated with large amplitude transverse waves. We conclude by noting that it may be interesting to investigate in-situ signatures of wave-particle resonances such as those reported in our hybrid simulation via the field-particle correlation technique~\citep{klein_2016}.

\section{Summary}
\label{summary}

We have considered dispersive and kinetic effects on a 2D Alfv\'en wave packet with constant magnetic field pressure similar to a switchback in a low-$\beta$ plasma. This complements our previous work where the parametric decay of a 2D switchback was investigated in MHD. Our results can be summarized as follows:

\begin{itemize}

\item{Dispersion due to the Hall term introduces a characteristic time $\tau^*=\tau_a \ell/d_i$, where $\ell$ is the wave packet's smallest scale (parallel, $\ell_x$, or transverse, $\ell_y$, to the guiding field), $\tau_a=\ell_x/v_a$ and $d_i$ the proton inertial length.}
\item{If $\ell/d_i\gg 1$, dispersion of the initial wave packet starts to affect its dynamics after a time $t\simeq\tau^*$.}
\item{If $\ell/d_i\gtrsim 1$ the wave packet couples with compressible and other dispersive modes at early times $t<\tau^*$. The  wave packet then  starts to disperse and disrupt at time $t\simeq\tau^*$. }
\item{In the Hall-MHD model, coupling with compressible modes leads to the gradual conversion of the wave packet's kinetic and magnetic energy into internal energy.}
\item{When proton kinetic effects are included,  the generation of dispersive waves and slow-modes is inhibited. Compressible fast modes are emitted by the wave packet, which undergo Landau resonance at the Alfv\'en speed leading to parallel heating.} 
\item{The resulting heating rates (parallel and total) are estimated to be of about the same order of the heating rates estimated from  fast solar wind observations at $R\simeq0.6$~au}.
\item{Observationally, we expect that the shortest switchbacks, of duration $\delta t\lesssim 100$~s,  display signatures of fast-modes and dispersive waves propagating from their leading and trailing edge, both along the radial and transverse directions.}
\end{itemize}

\section{Appendix}

The plots shown in Fig.~\ref{data} were obtained  by combining PSP data at radial distances $0.06<R< 0.5$~au (E1-E12) with data from Solar Orbiter at distances $0.5<R<1$~au (between June 1, 2018 to March 1, 2022). We have used Level 2 magnetic field measurements from the Flux Gate Magnetometer (FGM) \citep{bale_fields_2016} onboard PSP, as well as Level 3 plasma moment data from the Solar Probe Cup (SPC) for E1-E8, and Solar Probe Analyzer (SPAN) part of the Solar Wind Electron, Alpha and Proton (SWEAP) suite for E9-E12 \citep{kasper_solar_2016}. The plasma data are comprised of moments of the distribution function including the proton velocity vector $\boldsymbol{V}_{p}$, number density $n_p$, and temperature $T_p$. When available, electron number density data derived from the quasi-thermal noise from the FIELDS instrument \citep{Moncuquet_2020}, have been used for estimating proton number density. In order to get the proton density from the electron density, one must consider charge neutrality, and consequently a $\approx4\%$  abundance of alpha particles. Accordingly, electron density from QTN has been divided by 1.08. For Solar Orbiter data, we used magnetic field measurements from the Magnetometer (MAG) instrument \citep{horbury2020solar} and particle moments from the Proton and Alpha Particle Sensor (SWA-PAS) onboard the Solar Wind Analyser (SWA) suite of instruments \citep{owen_solar_2020}. 

Starting from this dataset, we have then  identified and removed heliospheric current sheet crossings and magnetic field data were resampled at 1 second resolution. We have then determined the duration $\delta t$ of field reversals. These are defined as those time intervals in which 
\begin{equation}
\theta_{sb}\equiv\arccos({\bf B\cdot <B>}/(|{\bf B}||<{\bf B}>|))>\pi/2,
\end{equation}

where here we have calculated the mean magnetic field $<{\bf B}>$ over a time interval $\Delta t=8$~h. We defer the reader to \citet{tenerani_2021} for  details on the identification process  of switchbacks. For each identified switchback, we have determined the field reversal duration $\delta t$, the  mean solar wind radial velocity $V_r$, $v_a$ and $d_i$.  Assuming that switchbacks are advected at the solar wind speed in the radial direction, we have then defined $\delta\ell=\delta t |V_{r}+v_a-V_{sc}|$ with $V_{sc}$ the spacecraft speed.

\acknowledgements{This research was supported by NASA grant \#80NSS\-C18K1211. We acknowledge the Texas Advanced Computing Center (TACC) at The University of Texas at Austin for providing HPC resources that have contributed to the research results reported within this paper. URL: http://www.tacc.utexas.edu.}

\end{document}